%% file: mobileng_2014.tex
\documentclass[runningheads,a4paper]{llncs}

\usepackage{amsmath,amssymb,amsfonts,latexsym}
\usepackage{graphicx}
\usepackage{times}
\usepackage{graphicx}
\usepackage{xspace}
\usepackage{color}
\usepackage{pifont}
\usepackage{eurosym}
\usepackage{listings}
\usepackage{url}
\usepackage{balance}
\usepackage{enumitem}
\usepackage{dsfont}
\usepackage{subfigure}
\usepackage{wrapfig}
\usepackage{listings}
\usepackage{comment}

\usepackage{epsfig}
\usepackage{enumerate}
\usepackage{epsf,picinpar}
\usepackage{varioref}
\usepackage{colortbl,multirow,hhline}
\usepackage{algorithmic}
\usepackage{algorithm}
\usepackage[normalem]{ulem} 
\usepackage{xcolor}
\usepackage{array}

\begin{document}

\mainmatter  

\title{Stakeholders, Viewpoints and Languages of a Modelling Framework for the Design and Development of Data-Intensive Mobile Apps}
\titlerunning{Stakeholders, Viewpoints and Languages of a Modelling Framework for the Design and Development of Data-Intensive Mobile Apps}

\author{Mirco Franzago$^1$, Ivano Malavolta$^2$, Henry Muccini$^1$}

\institute{$^2$Department of Information Engineering, Computer Science and Mathematics, \\ University of L'Aquila, Italy \\
$^2$Gran Sasso Science Institute, L'Aquila, Italy\\
mirco.franzago@univaq.it, ivano.malavolta@gssi.infn.it, henry.muccini@univaq.it}








\maketitle
\thispagestyle{empty}
\pagestyle{empty}

\begin{abstract}
Today millions of mobile apps are downloaded and used all over the world. Guidelines and best practices on how to design and develop mobile apps are being periodically released, mainly by mobile platform vendors  and researchers. They cover different concerns, and refer to different technical and non-technical stakeholders.
Still, mobile applications are developed with ad-hoc development processes, and on-paper best practices. 

In this paper we discuss a multi-view modelling framework supporting the collaborative design and development of mobile apps. The proposed framework embraces the Model-Driven Engineering methodology. 
%
This paper provides an overall view of the modelling framework in terms of its main stakeholders, viewpoints, and modelling languages.

%


\end{abstract}

\input{intro}
\input{motivation}
\input{approach}
\input{conclusion}

\bibliographystyle{splncs}
\balance
\bibliography{mobilesoft2014}
\end{document}

%% file: intro.tex
\section{Introduction}\label{sect:introduction}

 
Mobile devices are replacing traditional desktop and mobile instruments; 
people relies on mobile devices for surfing the web, purchasing products, or to be
part of the social network; the mobile applications market now counts more than two millions applications, downloaded billions of times per year.
As an indicator of this situation, today the total activity on smartphones and tablets accounts for an incredible 60\% of the time spent on digital media in the United States~\cite{comscore}.


Still, mobile applications are developed with ad-hoc development processes, and
on-paper best practices~\cite{Wasserman,realChallengesMobile}. 
Also, best practices for app design and development have been released by mobile platform vendors (e.g., \cite{AndroidDev}), by the W3C consortium, and by various practitioners and researchers. Still, what the community is missing is a systematic approach, and related automated tools, that may translate the known guidelines into automated tasks that mobile developers can run in order to develop quality apps.




In this paper we discuss a dedicated modelling framework for automating the design and development of data-intensive mobile applications, i.e., apps whose primary purpose is to present a large amount of content (as further discussed and motivated in Section \ref{sec:motivation}). With this scope in mind, our framework makes use of Model-Driven Engineering~\cite{MDE_2002} technologies for supporting the multi-view modelling and development of mobile applications. 
Through four main mobile-specific models (namely, the {\em Navigation}, {\em Data}, {\em UI}, and {\em Business Logic} models) the various stakeholders involved in the development of a mobile application can {\em collaboratively} work for designing the app. 
In this paper we describe the modelling framework in terms of its stakeholders, viewpoints, and modelling languages.
%
%


The rest of the paper is organized as follows. In Section \ref{sec:motivation} we describe the data-intensive mobile applications domain and their main design issues and challenges. Section \ref{sec:approach} presents our modelling framework, and Section~\ref{sect:conclusions} closes the paper.

%% file: motivation.tex
\section{Data-intensive mobile applications}\label{sec:motivation}
%

This research is concerned with enhancing the design and development activities
of \textit{data-intensive mobile applications}, i.e., those applications whose primary purpose is to present a large
amount of content to a variety of possible users. To further scope data-intensive apps, and by building on the definition of data-intensive web-sites \cite{ceri2003morgan}, a data-intensive mobile app differs from other mobile apps for their:
\begin{itemize}
	\item focus on browsing collections of data and basic interactions with data items, with simpler functional requirements;
	\item support of one-to-one content delivery, where each user must have the impression of interacting with an interface specifically tailored to her needs and preferences;
	\item simpler transactional requirements, in most cases limited to high-performance read-only access and standard write operations of a well-delimited fraction of the data;
	\item support for delivering content to multiple devices;
	\item focus on information organization and ease of navigation where users can immediately understand the structure of the mobile application.
\end{itemize}
%
%



Current practices in data-intensive mobile applications development are still plagued by a number of recurring issues and challenges~\cite{Wasserman}.
In order to have a clear understanding of those issues, we are working closely with industry partners
in the context of research projects (e.g., ~\cite{ehealth}).
Also, we are continuously performing informal interviews with developers\footnote{We are working with industry partners and one of the authors is a mobile applications developer with more than twenty projects in his portfolio~\cite{ivanoPortfolio}.}, projects managers and other involved stakeholders for completing our knowledge about this application domain.

Even if many of the identified issues have a technical nature (e.g., mobile devices fragmentation, multi-device testing,
code reuse across platforms), 
from our experience and interactions with professionals in the field, we actually noted that
issues and challenges pertaining to the \textbf{design} of a mobile application can have a much bigger and disastrous impact on its success~\cite{fling2009mobile,Wasserman}. 
In the following we report the main issues and challenges of designing data-intensive mobile apps:

\begin{itemize}
  \item partial reasoning in the context,
  \item limited information architecture engineering,
  \item stakeholders diversity.
\end{itemize}

A further elaboration of the aforementioned issues and challenges about the design of data-intensive mobile apps is provided in \cite{Mobilesoft2014}.

In order to overcome the above mentioned problems,
we carry out a top-down reasoning process with the chief aim of
specifying a design and development framework that can satisfy the needs of data-intensive mobile applications stakeholders.
In the next section we will present our framework.

%
%

%% file: approach.tex
\section{The modelling framework}\label{sec:approach}
By exploit the well-known principle of multi-viewpoint specification~\cite{MDE_2002}, our modelling framework allows stakeholders to represent a mobile app as a set of independent models and correspondences among them. The latter is an elegant solution for managing complexity since it allows each stakeholder to focus on those aspects of the app in which she is expert, abstracting away from the rest of the concerns.

In order to reason in an organized manner on our framework, we identified the set of stakeholders and viewpoints of a data-intensive mobile project.
Table 1 shows the results of our analysis.
Basically, we identified four main viewpoints of a mobile project: \textit{navigation} refers to the navigation flow and the logical structure of the mobile application,
\textit{data} refers to the structure and operations of the data managed by the application, \textit{UI} refers to the presentation, layout and graphical style of the application, and \textit{business logic} mainly refers to the internal behaviour of the application and the interaction between the user and the application user interface.

The stakeholders of a mobile project can be either technical (like back-end developer, UI designer) or non-technical (such as users, customers, or content producers).

\renewcommand{\arraystretch}{1.65}
\begin{table}[hb!]
\center
    \begin{tabular}{|p{3cm}|>{\centering\arraybackslash}m{2cm}|>{\centering\arraybackslash}m{2cm}|>{\centering\arraybackslash}m{2cm}|>{\centering\arraybackslash}m{2cm}|}
    \hline
    ~                     & Navigation & Data 		& UI 						 & Business logic				\\ \hline
    UI designer           & \checkmark & \checkmark	& \checkmark     & \checkmark           \\
    App developer         & \checkmark & \checkmark	& \checkmark     & \checkmark           \\
    Back-end developer    & ~          & \checkmark & ~              & \checkmark           \\
    Information architect & \checkmark & \checkmark & \checkmark     & ~           					\\
    Content producer      & ~          & \checkmark & ~              & ~           					\\
    User                  & ~					 & ~       		& \checkmark     & ~           					\\
    Customer              & \checkmark & ~       		& \checkmark     & ~           					\\
    Project manager       & \checkmark & ~       		& \checkmark     & ~           					\\ 		\hline
    \end{tabular}\label{tab:ciao}\caption{Stakeholders and viewpoints of the modelling framework}
\end{table}

In data-intensive mobile projects the whole implementation of the system can be split among three different stakeholders:
the \textit{app developer} who is in charge of implementing the mobile application with a focus on its client-side business logic, the \textit{back-end developer} who is in charge of implementing the server-side logic and interfaces to make the data available to the mobile application, and the \textit{UI designer} who is in charge of developing the user interface of the application, thus focussing on the presentation part of the project.
Also, the \textit{information architect} takes care of the design of the layout of the various views of the mobile application, reasons on its usability, navigability, etc.
\textit{Content producers} are in charge of creating the contents accordingly to the structure proposed by the information architect, developers, and UI designer.
\textit{Users} and \textit{customers} should refer to the user interface of the application (together with its prototypes) in order to assess the validity of the various choices taken by the other stakeholders.

A clear definition of the stakeholders and their viewpoints with respect to a data-intensive mobile app helps us in two main directions:
(i) it supports us in clearly reasoning on the actors developing a data-intensive mobile application and their concerns to be considered, and
(ii) it plays the role of blueprint for the concrete realization of our framework.

Our modeling framework is composed of four main languages: navigation, data, UI, and business logic, more specifically:

\begin{itemize}
	\item \textit{Navigation}. It represents the structure of a data-intensive mobile application in terms of its \textit{views} and the \textit{navigation flow} between them.
	According to our language, views are the atomic first-class elements of a navigation model, views are linked by means of navigation flows. A navigation flow can optionally
	contain a condition to decide if the flow can be undertaken by the user, depending on the current state of the mobile application.
	
	\item \textit{Data}. It focusses on the structure of the various data types underlying the mobile application under development.
	More specifically, a data model is used to describe (i) each data entity that is managed by the application, (ii) its properties, (iii) its exposed operations, and (iv) its references to other data entities.
	It is important to note that the data model language is platform-independent, it does not prescribe any constraint about how to manage persistence of the modelled data entities (e.g., local storage, in the cloud, etc.), nor the details about their operations.
	
	\item \textit{UI}. It describes the layout of each view of the mobile application, together with the UI elements composing it.
	UI elements can be either basic (e.g., button, label, map, image, navigation item, etc.) or containers (e.g., list items, grids, menus, navigation bars, etc.); clearly,
	containers can be composed of either basic or other container elements.
		
	\item \textit{Business Logic}. It describes the internal operations of the application and the interaction between the user and the mobile application interface.
	This language is based on the event-condition-action paradigm. Actions can refer to operations on data entities, to the update of an UI element, to the navigation through a specific navigation flow, to the
	access of a device capability (like GPS sensor or camera). Events can refer to user-generated interactions (e.g., a gesture on an UI element),
	to a specific device capability (e.g., disconnection of the device from	the network, a callback from a GPS sensor, etc.), or they can be application-specific (e.g., logout of the user, availability of a given data
	from a remote server, etc.).
	Actions can be executed in sequences and in response to the triggering of events, thus enabling a control flow between them.
	Optionally, conditions can be associated to control flows; the interaction flow goes through a specific sequence of actions and events only if its condition evaluates to true.
\end{itemize}

The main drivers we followed during the design of those language are: simplicity, platform independence, conciseness, and being self-contained.
The modelling languages are (loosely) coupled by means of typed correspondences
linking the various concepts across models in a non-intrusive manner.
Examples of correspondences include those linking:
a view in the navigation model to its corresponding main container in the UI model, an attribute of an entity in the data model to the label in the UI model showing its value to the user, an action in the interaction model to the corresponding CRUD operation in the data model to be performed when the action is executed, etc. Thanks to those inter-model correspondences, each model is self representative and can be re-used across projects and organizations.
%
%

%% file: conclusion.tex
\section{Conclusions}\label{sect:conclusions}

This paper presented the main stakeholders, viewpoints, and modelling languages of a collaborative modelling framework for mobile apps. Based on Model-Driven Engineering principles,
the framework supports the collaborative modelling of mobile applications through four main principal views, and with the aim of pre-deployment analysis and code generation.